\documentclass[conference,10pt]{IEEEtran}
\ifCLASSINFOpdf
\else
\fi
\ifCLASSINFOpdf
\else
\fi

\usepackage{array,booktabs}
\usepackage{algpseudocode}
\usepackage{algorithm}
\usepackage{amsfonts}
\usepackage{amsmath}
\usepackage{amssymb}
\usepackage{amsthm}
\usepackage{graphicx}
\usepackage[noadjust]{cite}
\usepackage{mathrsfs}
\usepackage{stmaryrd}
\usepackage{siunitx}
\usepackage{multicol}
\usepackage{tikz}
\usepackage{setspace}
\usetikzlibrary{shapes}

\usepackage{pstricks, pst-node, pst-plot, pst-circ}
\usepackage{moredefs}
\providelength{\AxesLineWidth}       
\providelength{\plotwidth}           
\providelength{\LineWidth}           
\providelength{\MarkerSize}          
\newrgbcolor{GridColor}{0.8 0.8 0.8}%
\newcommand{\pbp}{person-by-person }

\begin{document}

\title{A General Method for the Design of Tree Networks Under Communication Constraints}

\author{\IEEEauthorblockN{Alla Tarighati, and Joakim Jald{\'e}n}
\IEEEauthorblockA{ACCESS Linnaeus Centre, Department of Signal Processing, \\
KTH Royal Institute of Technology, Stockholm, Sweden\\
Email: \{allat, jalden\}@kth.se}
}

\maketitle

\begin{abstract}
We consider a distributed detection system with communication constraints, where several nodes are arranged in an arbitrary tree topology, under the assumption of conditionally independent observations. We propose a cyclic design procedure using the minimum expected error probability as a design criterion while adopting a \pbp methodology. We design each node jointly together with the fusion center, while other nodes are kept fixed, and show that the design of each node using the \pbp methodology is analogous to the design of a network with two nodes, a network which we refer to as the \emph{restricted} model. We further show how the parameters in the restricted model for the design of a node in the tree network can be found in a computationally efficient manner. The proposed numerical methodology can be applied for the design of nodes arranged in arbitrary tree topologies with arbitrary channel rates for the links between nodes and for a general $M$-ary hypothesis testing problem.
\end{abstract}

\IEEEpeerreviewmaketitle

\section{Introduction}\label{sec:intro}
We consider a distributed, or decentralized, hypothesis testing problem in a general tree network configured as a directed graph, where observations are made at spatially separated nodes. The root of the graph is the fusion center (or FC), and information from nodes propagate toward the FC. If the nodes are able to communicate all their data to the FC, there is no fundamental difference from the centralized case, where the classical solution is to use threshold tests on the likelihood ratios of the received data at the FC. However if there are communication constraints on the links between the nodes, the nodes need to carry out some processing and give a summarized, or quantized, version of their data as output.

The problem of optimal decentralized hypothesis testing has gained noticeable interest over the last 30 years, see for instance \cite{Tsi93,TanPK93,Pet94,Varsh96,Vis97} and references therein. A common goal in these references is to find a strategy which optimizes a performance measure, like minimizing the error probability at the FC. However, it is difficult to derive the optimal processing strategies at the nodes in distributed networks, even for small size networks. Therefore most of the works on this topic focus on \pbp optimization as a practical way for the design of decentralized networks. Using \pbp optimization, it is guaranteed that the overall performance at the FC is improved (or, at least not worsened) with every iteration of the algorithm. Unfortunately \pbp optimization gives a necessary, but in general not sufficient, condition for an optimal strategy \cite{Varsh96}.

While deriving decision function for one node in the \pbp methodology for the design of nodes in a general tree network (including parallel and tandem networks) all other nodes and the FC are assumed to have already been designed and remain fixed. Focusing on \pbp optimality, a typical result is that if observations at the nodes are independent conditioned on true hypothesis, likelihood ratio quantizers are \pbp optimal, while the optimal thresholds in the quantizers are given by the solution of systems of nonlinear equations, with as many variables as the number of thresholds. This however makes the computation of optimal thresholds intractable, even for a moderate size network \cite{Varsh96,TTW08a}.

The main contribution of this work is to show that -- contrary to previous claims -- it is possible to under the \pbp methodology design a distributed detection network arranged in an arbitrary tree topology with a reasonable computational burden. In order to do that we modify the \pbp methodology for the design of nodes in the tree topology by letting the FC update its decision function in every iteration together with the nodes. In other words, we adopt a \pbp methodology in which at every iteration each node is designed jointly together with the FC. We further assume that the FC uses the maximum a-posteriori (MAP) rule to make the final decision, which is motivated by the optimality of the MAP rule when the performance criterion is global error probability, or error probability at the FC.

In order to obtain a tractable solution during the design of a node, let say $m_0$ (together with the FC), all other nodes are modeled using a Markov chain and it will be shown that the design of $m_0$ is analogous to the design of a special case of a network with only two nodes (which is here called the \emph{restricted} model). Then we will show how the parameters for this restricted model can be found recursively in a computationally efficient way from the original network.

This paper is organized as follows. In Section \ref{sec:preliminaries} we present our model in detail and formulate the problem. In Section \ref{sec:restricted} we introduce the restricted model and describe how it can be obtained from the original tree network. We present numerical examples to illustrate the benefit of proposed approach in Section \ref{sec:examples} and Section \ref{sec:conclusion} concludes the paper.

\section{Preliminaries}\label{sec:preliminaries}

\begin{figure}[t]
\begin{center}
\begin{tikzpicture}[align=center,scale=0.7,>=stealth] 
\node (fc) at (0,0) [circle,minimum size=0.4cm,draw] {$\small\text{FC}$} ;
\node (n1) at (-2,1) [circle,minimum size=0.4cm,draw] {} ;\draw [->] (n1) -- (fc); \node  at (-1.4,1){\small$m_1$};
\node (n2) at (-3,-0.5) [circle,minimum size=0.4cm,draw] {} ;\draw [->] (n2) -- (n1); 
\node (n3) at (-3.5,1) [circle,minimum size=0.4cm,draw] {} ;\draw [->] (n3) -- (n1); 
\node (n4) at (-2,2.5) [circle,minimum size=0.4cm,draw] {} ;\draw [->] (n4) -- (n1); \node  at (-2.6,2.5){\small$m_2$};
\node (n5) at (-1,4) [circle,minimum size=0.4cm,draw] {} ;\draw [->] (n5) -- (n4); 
\node (n6) at (-3,4) [circle,minimum size=0.4cm,draw] {} ;\draw [->] (n6) -- (n4); \node  at (-3.6,4){\small$m_3$};
\node (n7) at (1,-2) [circle,minimum size=0.4cm,draw] {} ;\draw [->] (n7) -- (fc); \node  at (1.2,-1.6){\small$m_7$};
\node (n8) at (0,-3) [circle,minimum size=0.4cm,draw] {} ;\draw [->] (n8) -- (n7); \node  at (-0.6,-3){\small$m_8$};
\node (n9) at (2,-3) [circle,minimum size=0.4cm,draw] {} ;\draw [->] (n9) -- (n7); \node  at (1.4,-3){\small$m_9$};
\node (n10) at (3,-2) [circle,minimum size=0.4cm,draw] {} ;\draw [->] (n10) -- (n7);\node  at (3,-1.5){\small$m_{10}$};
\node (n11) at (2,2) [circle,minimum size=0.4cm,draw] {} ;\draw [->] (n11) -- (fc); \node  at (1.4,2){\small$m_4$};
\node (n12) at (3,1) [circle,minimum size=0.4cm,draw] {} ;\draw [->] (n12) -- (n11); 
\node (n13) at (1,3) [circle,minimum size=0.4cm,draw] {} ;\draw [->] (n13) -- (n11); 
\node (n14) at (4,3) [circle,minimum size=0.4cm,draw] {} ;\draw [->] (n14) -- (n11); \node  at (4.6,3){\small$m_5$};
\node (n15) at (5,2) [circle,minimum size=0.4cm,draw] {} ;\draw [->] (n15) -- (n14); 
\node (n16) at (3,4) [circle,minimum size=0.4cm,draw] {} ;\draw [->] (n16) -- (n14); 
\node (n17) at (5,4) [circle,minimum size=0.4cm,draw] {} ;\draw [->] (n17) -- (n14); \node  at (5.6,4){\small$m_6$};
\draw [->] (fc) -- (-1,-1); \node  at (-1.5,-1){$u_{N}$};
\end{tikzpicture}
\caption{An example network (observations are not shown).}
\label{fig:topol}
\end{center}
\end{figure}
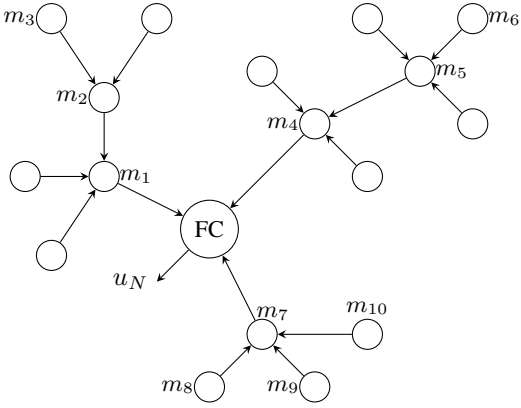
We describe a tree network by a directed and acyclic graph where the fusion center is the root of the network and information flows from every node on a unique path toward the root. We denote the tree network by $T \triangleq (V, E)$, where $V=\{m_1, m_2, \ldots, m_N\}$ is the set of $N$ nodes and $E=\{e_{i,j}\}$ is the set of directed edges from node $m_i$ to node $m_j$. Without loss of generality we assume that node $m_N$ is the fusion center (labelled FC in Fig.~\ref{fig:topol}). 

We say that node $m_i$ is the \emph{predecessor} of node $m_j$ if there is a directed path from node $m_i$ to node $m_j$, and say that node $m_j$ is a \emph{successor} of node $m_i$. Accordingly we say that node $m_i$ is an \emph{immediate} predecessor of node $m_j$ if $e_{i,j}\in E$ (equivalently node $m_j$ is the immediate successor of node $m_i$). This is exemplified in Fig.~\ref{fig:topol} where nodes $m_3$ and $m_2$ are predecessors of node $m_1$, while $m_2$ in an immediate predecessor of node $m_1$. The set of all immediate predecessors to node $m_i$ is denoted by $I_i$, and the set of all immediate predecessors to the fusion center is denoted by $I_{f}$. For instance, $I_f=\{m_1, m_4, m_7\}$ in Fig.~\ref{fig:topol}. We also define $S_i$ as a set consisting of node $m_i$ and all its successors, excluding the FC. In other words, $S_i$ is the set of all nodes the input messages of node $m_i$ pass through to reach the FC, e.g., $S_6=\{m_6, m_5, m_4\}$ in the example of Fig.~\ref{fig:topol}. We further define the \emph{last successor} $ls(m_i)$ of node $m_i$ as the last node that the input of node $m_i$ passes through before it reaches the FC, i.e., $ls(m_i)= m_l$, if $m_l \in S_i $ and $m_l\in I_f$. We let $T_i \triangleq (V_i,E_i)$ define the sub-tree of the network with node $m_i$ as its root, where $V_i$ and $E_i$ are the set of nodes and directed edges in sub-tree $T_i$.

We assume that there are two types of nodes in the tree network: \emph{leaves} and \emph{relays}. A leaf is a node which makes observation and a relay is a node that only receives messages from its immediate predecessors. In Fig.~\ref{fig:topol} node $m_3$ and $m_2$ exemplify a leaf and a relay, respectively. Without loss of generality nodes which both make an observation and receive messages from their immediate predecessors are considered to be relays, since every observation can equivalently be considered as the output of a leaf with output cardinality equal to the cardinality of its observation space. Let $C_l$ be the set of all leaves and $C_r$ be the set of all relays in the network. In a tree network, each leaf $m_i\in C_l$ using its own observation $x_i\in \mathcal{X}_i$ makes a decision $u_i\in \mathcal{M}_i$ and sends it through a rate-constrained channel ($e_{i,j}$) to its immediate successor $m_j$ (which is a relay or FC). Each relay $m_i\in C_r$, using input messages from all of its immediate predecessors $I_{i}$, makes a decision $u_i\in \mathcal{M}_i$ and sends it through a rate-constrained channel to its immediate successor. Eventually the fusion center makes the final decision $u_N$ from the set $\{0, 1, \ldots, M-1\}$ in an $M$-ary hypothesis testing problem. In this paper, we restrict our attention to discrete observation spaces $\mathcal{X}_i$ where $m_i\in C_l$. However we wish to stress that any continuous observation space can be approximated by a discrete observation space, by representing the continuous space by a set of intervals indexed by $x_i$ from the discrete space \cite{LLG90,Alla14}.

The channel between node $m_i$ and its successor is considered to be an error-free but rate-constrained channel with rate $R_{i}$ bits. The output of node $m_i$ is then from a discrete set $\mathcal{M}_i$ with cardinality $\Vert \mathcal{M}_i\Vert =2^{R_{i}}$. Without loss of generality we assume that the output of node $m_i$ is from the discrete set $\mathcal{M}_i=\{1, \ldots, 2^{R_i}\}$. In this setup each node is a scalar quantizer which maps its inputs to an output message using a decision function $\gamma_i$. A leaf $m_l$ maps its observation $x_l$ to an output message $u_l$ using the function $\gamma_l:\mathcal{X}_l\to\mathcal{M}_l$, i.e.,
\begin{equation*}
\gamma_l(x_l)=u_l,
\end{equation*}
whereas a relay $m_r$ maps its input vector containing messages from its immediate predecessors $\{u_i:m_i\in I_r\}$ to an output message $u_r$ using the function $\gamma_r: \mathcal{M}_{I_r}\to\mathcal{M}_r$, i.e.,
\begin{equation*}
\gamma_r(\{u_i:m_i\in I_r\})=u_r.
\end{equation*}
$\mathcal{M}_{I_r}$ is defined as the product of alphabet of immediate predecessors to node $m_r$. For example, relay $m_7$ in Fig.~\ref{fig:topol} has three immediate predecessors, $I_7=\{m_8, m_9, m_{10}\}$, with decision spaces $\mathcal{M}_8$, $\mathcal{M}_9$ and $\mathcal{M}_{10}$, respectively, and $\mathcal{M}_{I_7}=\mathcal{M}_{8}\times\mathcal{M}_{9}\times\mathcal{M}_{10}$. We will use the terminology ``message" and ``index" interchangeably to denote the output of a node in the network.

We assume that the observations at the leaves, conditioned on the hypotheses, are independent. Then acyclicity of the network implies that the inputs to each relay, and also to the FC, are conditionally independent. We further assume that the conditional probability masses of the observations are known and denoted $P_j(x_l)\triangleq P(x_l\vert H_j)$, \, $j=0, 1, \ldots, M-1$, for an $M$-ary hypothesis testing problem.

In this paper, the objective is to arrive at a simple method for the design of the nodes decision functions, $\gamma_1, \ldots, \gamma_N$, in the tree network, in such a way that the global error probability at the FC is minimized. In order to derive the decision function at a node, we use the \pbp methodology and assume that all other nodes have already been designed and remain fixed. However, in contrast to previous works, we treat the FC in a different way than the other nodes: the FC decision function $\gamma_N$ is always updated together with the node decision function $\gamma_l$ currently being optimized.

If the decision functions of all the leaves and all the relays are fixed, the optimal fusion center will use the maximum a-posteriori (MAP) rule in order to make the global decision $u_N$ in favor of one of the hypotheses. Defining ${{u}}_f$ as the vector containing the messages from the immediate predecessors of the FC, i.e., ${{u}}_f\triangleq\{u_i:m_i\in I_{f}\}$, the FC  decides on the hypothesis $H_{\hat{m}}$ if \cite{Lap09}
\begin{equation}
\pi_{\hat{m}} P({{u}}_f\vert H_{\hat{m}})=\max_{j}\big\{\pi_{j}P({{u}}_f\vert H_j)\big\}\,,
\label{eq:FCrule}
\end{equation}
where $\pi_{j}\triangleq P(H_{j})$ is the a-prior probability of hypothesis $H_j$, and where $j \in \{0, 1, \ldots, M-1\}$ for the $M$-ary hypothesis testing problem. The expected minimum error probability in estimating $H$ given an input message vector ${{u}}_f$ from the set ${\displaystyle\mathcal{M}_{I_{f}}\triangleq \prod_{m_i\in I_{f}} \mathcal{M}_i}$ is \cite{Fed94}
\begin{equation}
P_\mathrm{E}=1-\sum_{{{u}}_f\in \mathcal{M}_{I_{f}}} \max_{j} \big\{\pi_{j}P({{u}}_f\vert H_{j})\big\}\,.
\label{eq:Pe}
\end{equation}
Knowing the conditional probabilities of the input messages to the FC from its immediate predecessors, the error probability $P_{\mathrm{E}}$ at the fusion center can be uniquely computed.
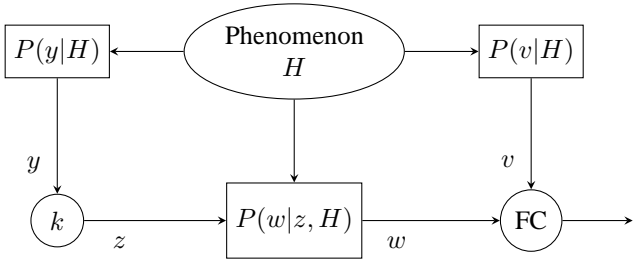
\begin{figure}[t]
\begin{center}
\begin{tikzpicture}[align=center,scale=0.9,>=stealth] 
\node (nl) at (-1.5,3) [circle,minimum size=.7cm,draw] {$k$};
\node (ch) at (2,3) [rectangle,minimum size=1.0cm,draw] {$P(w\vert z,H)$};
\node (fc) at (5.5,3) [circle,minimum size=.7cm,draw] {FC};
\node (a) at (-1.5,5.5) [rectangle,minimum size=.7cm,draw] {$P(y\vert H)$};
\node (b) at (5.5,5.5) [rectangle,minimum size=.7cm,draw] {$P(v\vert H)$};
\node (PH) at (2,5.5) [ellipse,minimum size=0.95cm,draw] {Phenomenon\\$H$};
\draw [->] (nl.east) -- (ch) node  [near start,below,inner sep=6pt] {$z$};
\draw [->] (ch.east) -- (fc) node [near start,below,inner sep=6pt] {$w$};
\draw [->] (fc.east) --(7,3)   node [near end,below,inner sep=6pt] {};
\draw [->] (PH) -- (a) node [near end,left,inner sep=6pt] {};
\draw [->] (a) -- (nl.north) node [near end,left,inner sep=6pt] {$y$};
\draw [->] (PH) -- (b) node [near end,left,inner sep=6pt] {};
\draw [->] (b) -- (fc) node [near end,left,inner sep=6pt] {$v$};
\draw [->] (PH) -- (ch) node [near end,right,inner sep=6pt] {};
\end{tikzpicture}
\caption{Restricted model for the design of nodes in tree topology.}
\label{fig:restricted}
\end{center}
\end{figure}

In the next section we will show that the design of a node in the network is analogous to the design of a node (labeled by $k$) in the restricted model as shown in Fig.~\ref{fig:restricted}, where the FC in both networks use the MAP rule \eqref{eq:FCrule} as the fusion decision function. We will further show how the conditional probabilities in the restricted model can be recursively computed from the original tree network.

\section{Restricted Model}\label{sec:restricted}
Consider the distributed network with two nodes, $k$ and FC, illustrated in Fig.~\ref{fig:restricted}. The fusion center FC using its input messages $w$ and $v$ makes a decision according to the MAP rule \eqref{eq:FCrule}. Let $w$ and $v$ be from the discrete sets $\mathcal{M}_w$ and $\mathcal{M}_v$, respectively. Conditioned on hypothesis $H_j$, the input messages $y$ and $v$ are independent with known conditional probability masses $P_j(y)$ and $P_j(v)$, respectively. Node $k$ maps its input $y$ from a discrete set $\mathcal{M}_y$ to an output $z$ from a discrete set $\mathcal{M}_z$ according to a decision function $\gamma_k:\mathcal{M}_y\to \mathcal{M}_z$, i.e., $\gamma_k(y)=z$. The index $z$ then passes through a discrete channel which maps it to the index $w$ with a known transition probability $P(w\vert z,H)$ that depends on the present hypothesis $H$. $P(y\vert H)$ and $P(v\vert H)$ are probabilistic mapping from the observation space to the discrete sets $\mathcal{M}_y$ and $\mathcal{M}_v$, respectively.

We show next that under the \pbp methodology, the design of a node, let say $m_0$, in an arbitrary tree network is analogous to the design of node $k$ in the restricted model for a particular instance of the parameters of the restricted model. To see this, let
\begin{equation}
        y \triangleq \left\{ \begin{array}{ll}
        x_0 & \mbox{if $m_0\in C_l$}\\
          \{u_i:m_i\in I_{0}\}  & \mbox{if $m_0\in C_r$}\, ,\end{array} \right.
\label{eq:udef} \end{equation}
be the complete input of node $m_0$ in the original network and let
\begin{equation}v\triangleq\{u_i:m_i\in I_{f}, \, m_i\neq ls(m_0)\},\label{eq:vdef}\end{equation}
be the complete input of the FC from its immediate predecessors $I_{f}$, excluding the node that has $m_0$ as its predecessor (the immediate predecessor of FC whose the path from $m_0$ goes through it to reach the FC), and assume that the fusion center in the restricted model uses the MAP rule \eqref{eq:FCrule}. The conditional PMFs of the inputs to node $k$ and FC in the restricted model are, due to the independency of observations and acyclicity of the network, then given by
\begin{equation} P_j(y) = \left\{ \begin{array}{ll}
          P_j(x_0) & \mbox{if $m_0\in C_l$}\vspace{2mm}\\
{\displaystyle\prod_{u_i: m_i\in I_{0}}}P_j(u_i)  & \mbox{if $m_0\in C_r$}\,, \end{array}\right.
\label{eq:Pudef} \end{equation}
and
\begin{equation}
P_j(v)=\prod_{\substack{u_i:m_i\in I_{f}\\ m_i\neq ls(m_0)}}P_j(u_i)\,.\label{eq:Pv}
\end{equation}
The transition probability $P(w\vert z,H_j)$ is then simply the transition probability from $u_0$ to $u_K$, where $u_K$ is the output message of the last successor of $m_0$, i.e., $m_K\triangleq ls(m_0)$. The key point is that under the \pbp methodology for the design of node $m_0$ together with the FC, all other nodes in the network remain fixed. This implies that $P_j(v)$, $P_j(y)$ and $P(w\vert z, H_j)$ remain fixed and together with the structure of the restricted model, they capture all the important aspects of the joint design problem of $m_0$ and the FC. In the rest of this section we will show how the transition probabilities $P(w\vert z, H_j)$ in the restricted model can be found from the original network and describe a recursive method for the computation of the conditional probability masses $P_j(y)$ and $P_j(v)$ in the restricted model from the original tree network.

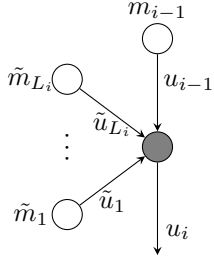
\begin{figure}
\begin{center}
\begin{tikzpicture}[align=center,scale=1.2,>=stealth] 
\node (mi) at (0,0) [circle,minimum size=.4cm,draw,fill=gray] {}; \node at (0,1.5) {$m_{i-1}$};
\node (m1i) at (0,1.2) [circle,minimum size=.4cm,draw] {}; 
\node (nl) at (-1,.75) [circle,minimum size=.4cm,draw] {}; \node at (-1.4,0.75) {$\tilde{m}_{L_i}$};
\node (n1) at (-1,-.75) [circle,minimum size=.4cm,draw] {}; \node at (-1.4,-0.75) {$\tilde{m}_1$};
\draw [->] (nl) -- (mi) node  [near start,below,inner sep=6pt] {};\node at (-0.5,0.25) {$\tilde{u}_{L_i}$};
\draw [->] (n1) -- (mi) node [near start,below,inner sep=6pt] {};\node at (-0.5,-0.6) {$\tilde{u}_1$};
\draw [->] (m1i) --(mi)   node [near start,right,inner sep=6pt] {}; \node at (.35,0.7) {$u_{{i-1}}$};
\draw [->] (mi) -- (0,-1.2) node [near end,right,inner sep=3pt] {$u_{{i}}$};
\node[rotate=90] at (-1,0) {$\ldots$};
\end{tikzpicture}
\end{center}
\caption{Node $m_i$ (shaded circle) and its ${L_i}+1$ immediate predecessors, $I_i$.}
\label{fig:mi}
\end{figure}
First, consider an arbitrary node in a tree network, say $m_i$, and assume that this node has $\vert I_i\vert=L_i+1$ immediate predecessors, containing node $m_{i-1}$. With a slight abuse of notation we define the set of immediate predecessors of node $m_i$ as $I_{i}\triangleq\{\tilde{m}_1, \ldots, \tilde{m}_{L_i}, m_{i-1}\}$, as exemplified in Fig.~\ref{fig:mi}. We also define $\tilde{u}_l\in\tilde{\mathcal{M}}_l$ as the output message of node $\tilde{m}_l,\, l=1, \ldots, L_i$. Conditioned on $H_j$, each index $u_{{i-1}}\in\mathcal{M}_{i-1}$ at the input of node $m_i$ is mapped to output index $u_{i}\in\mathcal{M}_i$ according to function
\begin{equation*}\gamma_{i}(\tilde{u}_1, \ldots, \tilde{u}_{L_i}, u_{i-1})=u_i\, ,\end{equation*}
with probability $P_j(u_{i}\vert u_{{i-1}})\triangleq P(u_{i}\vert u_{{i-1}}, H_j)$ which is equal to
\begin{equation}
\begin{split}
P_j(u_{i}\vert u_{i-1})
&=P_j\left(\gamma_{i}(\tilde{u}_1, \ldots, \tilde{u}_{L_i}, u_{i-1})\vert u_{i-1}\right)\\
&=\sum_{(\tilde{u}_1, \ldots, \tilde{u}_{L_i}) \in \gamma_i^{-1}(u_{i-1} ,u_{i})}\hspace{-5mm}P_j(\tilde{u}_1, \ldots, \tilde{u}_{L_i})\\
&=\sum_{(\tilde{u}_1, \ldots, \tilde{u}_{L_i}) \in \gamma_i^{-1}(u_{i-1} ,u_{i})}\hspace{-5mm}P_j(\tilde{u}_1) \ldots P_j(\tilde{u}_{L_i})\, ,\\
\label{eq:inversefun}
\end{split}
\end{equation}
where $P_j(\tilde{u}_{l})\triangleq P(\tilde{u}_{l}\vert H_j)$ is the conditional PMF of $\tilde{u}_{l}$, and where $\gamma_i^{-1}(u_{i-1}, u_{i})$ is the set of all input messages $(\tilde{u}_1, \ldots, \tilde{u}_{L_i})$ that satisfy $\gamma_{i}(\tilde{u}_1, \ldots, \tilde{u}_{L_i}, u_{i-1})=u_i$. It should be mentioned that, conditioned on the hypothesis, the input messages to node $m_i$ are independent. Now we can state the first important result from \eqref{eq:inversefun} as following: consider node $m_i$ and the set of its immediate predecessors, $I_{i}=\{\tilde{m}_1, \ldots, \tilde{m}_{L_i}, m_{i-1}\}$. Node $m_i$ has a Markovian behavior in the sense that, conditioned on the hypothesis and the input message $u_{i-1}$, its output message $u_i$ depends only on the inputs from the immediate predecessors $\{\tilde{m}_1, \ldots, \tilde{m}_{L_i}\}$ and not the sequence of messages preceding $m_{i-1}$. The transition probabilities for this Markov chain is found using \eqref{eq:inversefun}. The transition probability matrix from input index $u_{i-1}$ to output index $u_i$, conditioned on hypothesis $H_j$, is denoted ${\bf{P}}_j^{e_{i-1, i}}$ where the superscript $e_{i-1,i}$ indicates the specific edge that corresponds to the desired input $u_{i-1}$. ${\bf{P}}_j^{e_{i-1, i}}$ is with size $\Vert\mathcal{M}_i\Vert\times \Vert\mathcal{M}_{i-1}\Vert$, and its $(m, n)$th entry is defined as \cite{nor98} \begin{equation*}
{\bf{P}}_j^{e_{i-1, i}}(m, n)\triangleq P_j(u_{i}=m\vert u_{i-1}=n)\,.
\end{equation*}
Consequently, each relay node $m_i$ can be represented by $M\vert I_i\vert$ transition probability matrices ${\bf{P}}_j^{e_{k, i}}$, where $m_k\in I_i,$ and $j=0, \ldots, M-1$.

Consider again node $m_0$ and set $S_0=\{m_0, m_1, \ldots, m_K\}$. There is exactly one directed path from $m_0$ to the FC. Assume that node $m_1$ is the immediate successor of node $m_0$ and that node $m_{l+1}$ is the immediate successor of node $m_l$ for $l=1, \ldots, K-1$. The FC is the immediate successor of node $m_K$. After passing through $m_1$ to $m_K$, the output message $u_{0}$ of node $m_0$ from the discrete set $\mathcal{M}_0$ is mapped to an output message $u_{K}$ of node $m_K$ from the discrete set $\mathcal{M}_K$, which is then used as an input to the FC. The Markov property implies that the transition probability from $u_0$ to $u_{K}$ is given by
\begin{equation}\begin{split}
P_j(u_{K}\vert u_0)
&=\sum_{u_{1}}\ldots \sum_{u_{K-1}}P_j(u_{K}, u_{K-1}, \ldots, u_{1}\vert u_0)\\
&=\sum_{u_{1}}\ldots \sum_{u_{K-1}}\prod_{i=1}^{K}P_j(u_{i}\vert u_{i-1}, \ldots, u_0)\\
&\stackrel{}{=}\sum_{u_{1}}\ldots \sum_{u_{K-1}}\prod_{i=1}^{K}P_j(u_{i}\vert u_{i-1})\,.
\label{eq:markov1}
\end{split}\end{equation}
Equivalently, in matrix form if we define ${\bf{P}}^{0\to K}_j(m, n)\triangleq P_j(u_{K}=m\vert u_{0}=n)$, then \eqref{eq:markov1} implies
\begin{equation}
{\bf{P}}^{0\to K}_j={\bf{P}}^{e_{K-1,K}}_j \times \ldots \times {\bf{P}}^{e_{1,2}}_j \times {\bf{P}}^{e_{0,1}}_j.
\label{eq:markovmul}
\end{equation}
As there is only one directed path from node $m_0$ to $m_K$, we omitted the corresponding edge labels in ${\bf{P}}^{0\to K}_j$. Thus, using \eqref{eq:markovmul} we can replace all nodes between $m_0$ and the FC by a single hypothesis dependent transition probability given by ${\bf{P}}^{0\to K}_j$, when designing $m_0$. During the design of node $k$ in the restricted model (which is equivalent to the design of node $m_0$ in actual network) every channel transition probability $P_j(w=m\vert z=n)$ is replaced by the corresponding $(m,n)$th entry of ${\bf{P}}^{0\to K}_j$.

In forming the restricted model for the design of $m_0$ in the original network, in addition to channel transition probabilities $P(w\vert z, H)$, the transition probabilities $P(y\vert H)$ and $P(v\vert H)$ should be also determined. The input $y$ to the node $k$ is the complete input messages to node $m_0$ in the original tree. If node $m_0$ is a leaf then it only makes observation and $y=x_0$. However, if node $m_0$ is a relay, then $y$ is a vector containing input messages from its immediate predecessors according to \eqref{eq:udef} and $P(y\vert H)$ is defined as \eqref{eq:Pudef}. In the following we will show how $P_j(u_i)$ in a tree network [corresponding to $P_j(y)$ in the restricted model] can be found in a recursive manner. To this end, consider again node $m_0$ in the original tree network and its immediate predecessors $m_i\in I_0$. Suppose that node $m_i\in I_0$ receives messages from its $L_i$ immediate predecessors $I_i\triangleq\{\hat{m}_1, \ldots, \hat{m}_{L_i}\}$ and maps its input vector (denoted by $(\hat{u}_1, \ldots, \hat{u}_{L_i})$) to an output message $u_i$ according to a decision function $\gamma_i: \hat{\mathcal{M}}_1 \times \ldots \times \hat{\mathcal{M}}_{L_i}\to \mathcal{M}_i$, i.e., \begin{equation*}\gamma_i(\hat{u}_1, \ldots, \hat{u}_{L_i})=u_i\,.\end{equation*}
Then the probability masses $P_j(u_i)$ at the output of node $m_i$ are given by
\begin{equation}\begin{split}
P_j(u_i)
&=P_j\left(\gamma_i(\hat{u}_1, \ldots, \hat{u}_{Li})\right)\\
&=\sum_{(\hat{u}_1, \ldots, \hat{u}_{L_i})\in \gamma_i^{-1}(u_i)}\hspace{-5mm}P_j(\hat{u}_1, \ldots, \hat{u}_{L_i})\\
&=\sum_{(\hat{u}_1, \ldots, \hat{u}_{L_i})\in \gamma_i^{-1}(u_i)}\hspace{-5mm}P_j(\hat{u}_1) \ldots  P_j(\hat{u}_{L_i})\,,
\label{eq:Pui}\end{split}\end{equation}
where $\gamma_i^{-1}({u}_i)$ is the set of all input vectors $(\hat{u}_1, \ldots, \hat{u}_{L_i})$ that satisfy $\gamma_{i}(\hat{u}_1, \ldots, \hat{u}_{L_i})=u_i$. The last equation is the result of the fact that the inputs to each node in the tree network, conditioned on the hypothesis, are independent.

Equation \eqref{eq:Pui} shows how the probability masses of the output of node $u_i$ can be found based on the probability masses of its inputs and its decision function $\gamma_i$. Consider sub-tree $T_0$ in the network with node $m_0$ as its root. In the \pbp methodology used for the design of node $m_0$ we assume that all other nodes (except the FC) are kept fixed, including all the predecessors of node $m_0$ in its sub-tree $T_0$. Starting from the immediate predecessors of $m_0$ and going backward in the sub-tree, the probability masses of the output of each node can consequently be found based on the probability masses of its input and its decision function [cf.~\eqref{eq:Pui}]. Eventually, for a leaf $m_l$ in $T_0$ the probability masses at the output are given by
\begin{equation}
P_j(u_l)=\sum_{x_l\in \gamma^{-1}_l(u_l)}P_j(x_l) \, .
\label{eq:Pul}\end{equation}
Thus, the required PMFs at $m_0$ (represented by $P_j(y)$ in the restricted model) can be found by going forward from the leaves in $T_0$ toward node $m_0$. 
Using the same approach, $P_j(v)$ in the restricted model can be found in a recursive manner.

The minimum error probability at the FC in the restricted model (Fig.~\ref{fig:restricted}) is a function of the parameters $\gamma_k$, $P(y\vert H)$, $P(v,\vert H)$ and $P(w\vert z, H)$, i.e.,
\begin{equation}
P_{\mathrm{E,min}}=\mathcal{F}\big(\gamma_k, P(y\vert H), P(v,\vert H), P(w\vert z, H)\big)\,.
\end{equation}
Once the parameters in the restricted model are found, the error probability at the FC, when using the MAP criterion, is given by
\begin{equation}
P_\mathrm{E}=1-\sum_{v}\sum_{w} \max_{j} \big\{\pi_{j}P_j(v)P_j(w)\big\}\,, \label{eq:PeRest}
\end{equation}
where $P_j(w)\triangleq P(w\vert H_j)$ is
\begin{equation}\begin{split}
P_j(w)
&=\sum_{z\in \mathcal{M}_z}P_j(z)P_j(w\vert z)\\
&=\sum_{z\in \mathcal{M}_z}\,\sum_{y\in \gamma_k^{-1}(z)}P_j(y)P_j(w\vert z)\,,\label{eq:Pw}
\end{split}\end{equation}
and where $\gamma_k^{-1}(z)$ is the set of all input messages (vectors) $y$ that satisfiy $\gamma_{k}(y)=z$. Equations \eqref{eq:PeRest} and \eqref{eq:Pw} show how the error probability at the FC is affected by the parameters in the restricted channel (especially $\gamma_k$).

The goal of this paper is however not to show how $\gamma_k$ can be designed (together with the FC) in the restricted model. Rather, the goal is just to show how the optimization problem for $\gamma_k$ can be formulated compactly. However, in \cite{Alla14Tsp} a clear-cut guideline for the design of $\gamma_k$ with a reasonable computational burden is proposed. It is in \cite{Alla14Tsp} shown that the design of node $k$ in the restricted model can also be done in a \pbp manner in terms of the input set; an output index $z$ is assigned to a specific input $y$, while the assigned indices to other inputs are fixed.
\begin{figure}[t]
\begin{center}
\begin{tikzpicture}[align=center,scale=0.7,>=stealth] 
\node (fc) at (0,0) [circle,minimum size=0.4cm,draw] {{\small\text{FC}}} ;
\node (r1) at (-2,1.5) [circle,minimum size=0.4cm,draw] {{\small$r_1$}} ;\draw [->] (r1) -- (fc);
\node (l1) at (-3.25,3.25) [circle,minimum size=0.4cm,draw] {{\small$l_1$}} ;\draw [->] (l1) -- (r1);
\node (l2) at (-.75,3.25) [circle,minimum size=0.4cm,draw] {{\small$l_2$}} ;\draw [->] (l2) -- (r1);
\node (r2) at (2,1.5) [circle,minimum size=0.4cm,draw] {{\small$r_2$}} ;\draw [->] (r2) -- (fc);
\node (l3) at (3.25,3.25) [circle,minimum size=0.4cm,draw] {{\small$l_4$}} ;\draw [->] (l3) -- (r2);
\node (l4) at (0.75,3.25) [circle,minimum size=0.4cm,draw] {{\small$l_3$}} ;\draw [->] (l4) -- (r2);
\draw [->] (fc) -- (0,-1.5); \node  at (-.5,-1.25){{\small$u_N$}};
\end{tikzpicture}
\caption{$2$-symmetric $2$-uniform tree network.}
\label{fig:22tree}
\end{center}
\end{figure}
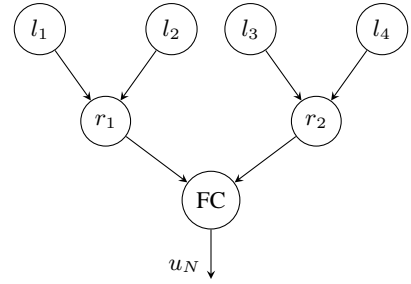

In closing, we emphasize that the proposed method for the design of nodes in the general tree topology (like other methods which use \pbp methodology) leads only to locally optimum solutions, which also depends on the initialization of the nodes. However, in the next section we show by a numerical example that good performance can nevertheless be obtained.
\section{Examples}\label{sec:examples}
In what follows we present some results from the application of the proposed method in the design of a tree network. We will consider a $2$-symmetric $2$-uniform tree network, as defined in \cite{TTW09} given in Fig.~\ref{fig:22tree}. The primary reason for choosing such a simple network is to be able to assess the performance of the proposed method through comparison with previous results for tree network and for specific channel rates. We assume the leaves $l_1, \ldots, l_4$ make observations $x_1, \ldots, x_4$, respectively, and the relays $r_1, r_2$ summarize the messages received from their corresponding immediate predecessors and the FC makes the final decision $u_N$ in favor of one hypothesis. We consider the case of binary hypothesis testing $M=2$, where real valued observations are, conditioned on the hypothesis, independent and identically distributed. The observation model at each leaf, where each observation consists of an antipodal signal $\pm a$ in unit-variance additive white Gaussian noise $n_i, i=1, \ldots, 4$, is given by
\begin{equation*}
\begin{split}
&H_0: {x_i}={-a} +{n_i} \\
&H_1: {x_i}={+a} + {n_i}\,.
\end{split}
\end{equation*}
The per channel signal-to-noise ratio (SNR) is then defined as $\mathcal{E}=\vert a \vert ^2$. We further assume equally likely hypotheses ($\pi_0=\pi_1=0.5$). Channels between the nodes are considered error-free but rate-constrained where the rate of the leaf-to-relay links are equal to $R_l$ bits, and the rate of the relay-to-FC links are equal to $R_r$ bits. This implies that the leaves' output massages are from the set $\{1, \ldots, 2^{R_l}\}$ and the relays' output messages are from the set $\{1, \ldots, 2^{R_r}\}$. The FC using the MAP rule \eqref{eq:FCrule} makes final decision $u_N$ from the set $\{H_0, H_1\}$.

In our simulations we initialized the relays with random functions, while for $R_l=1$ we initialized the leaves in all methods with the optimal local decision functions. For $R_l>1$ we uniformly quantized the two decision regions of the $R_l=1$ initialization.

\begin{figure}[t]
%
\psset{xunit=0.100000\plotwidth,yunit=0.225346\plotwidth}%
\begin{pspicture}(-6.198157,-4.388889)(5.115207,-0.418129)%


\psline[linewidth=\AxesLineWidth,linecolor=GridColor](-5.000000,-4.000000)(-5.000000,-3.946748)
\psline[linewidth=\AxesLineWidth,linecolor=GridColor](-4.000000,-4.000000)(-4.000000,-3.946748)
\psline[linewidth=\AxesLineWidth,linecolor=GridColor](-3.000000,-4.000000)(-3.000000,-3.946748)
\psline[linewidth=\AxesLineWidth,linecolor=GridColor](-2.000000,-4.000000)(-2.000000,-3.946748)
\psline[linewidth=\AxesLineWidth,linecolor=GridColor](-1.000000,-4.000000)(-1.000000,-3.946748)
\psline[linewidth=\AxesLineWidth,linecolor=GridColor](0.000000,-4.000000)(0.000000,-3.946748)
\psline[linewidth=\AxesLineWidth,linecolor=GridColor](1.000000,-4.000000)(1.000000,-3.946748)
\psline[linewidth=\AxesLineWidth,linecolor=GridColor](2.000000,-4.000000)(2.000000,-3.946748)
\psline[linewidth=\AxesLineWidth,linecolor=GridColor](3.000000,-4.000000)(3.000000,-3.946748)
\psline[linewidth=\AxesLineWidth,linecolor=GridColor](4.000000,-4.000000)(4.000000,-3.946748)
\psline[linewidth=\AxesLineWidth,linecolor=GridColor](5.000000,-4.000000)(5.000000,-3.946748)
\psline[linewidth=\AxesLineWidth,linecolor=GridColor](-5.000000,-4.000000)(-4.880000,-4.000000)
\psline[linewidth=\AxesLineWidth,linecolor=GridColor](-5.000000,-3.500000)(-4.880000,-3.500000)
\psline[linewidth=\AxesLineWidth,linecolor=GridColor](-5.000000,-3.000000)(-4.880000,-3.000000)
\psline[linewidth=\AxesLineWidth,linecolor=GridColor](-5.000000,-2.500000)(-4.880000,-2.500000)
\psline[linewidth=\AxesLineWidth,linecolor=GridColor](-5.000000,-2.000000)(-4.880000,-2.000000)
\psline[linewidth=\AxesLineWidth,linecolor=GridColor](-5.000000,-1.500000)(-4.880000,-1.500000)
\psline[linewidth=\AxesLineWidth,linecolor=GridColor](-5.000000,-1.000000)(-4.880000,-1.000000)
\psline[linewidth=\AxesLineWidth,linecolor=GridColor](-5.000000,-0.500000)(-4.880000,-0.500000)

{ \footnotesize 
\rput[t](-5.000000,-4.053252){$-5$}
\rput[t](-4.000000,-4.053252){$-4$}
\rput[t](-3.000000,-4.053252){$-3$}
\rput[t](-2.000000,-4.053252){$-2$}
\rput[t](-1.000000,-4.053252){$-1$}
\rput[t](0.000000,-4.053252){$0$}
\rput[t](1.000000,-4.053252){$1$}
\rput[t](2.000000,-4.053252){$2$}
\rput[t](3.000000,-4.053252){$3$}
\rput[t](4.000000,-4.053252){$4$}
\rput[t](5.000000,-4.053252){$5$}
\rput[r](-5.120000,-4.000000){$-4$}
\rput[r](-5.120000,-3.500000){$-3.5$}
\rput[r](-5.120000,-3.000000){$-3$}
\rput[r](-5.120000,-2.500000){$-2.5$}
\rput[r](-5.120000,-2.000000){$-2$}
\rput[r](-5.120000,-1.500000){$-1.5$}
\rput[r](-5.120000,-1.000000){$-1$}
\rput[r](-5.120000,-0.500000){$-0.5$}
} 

\psframe[linewidth=\AxesLineWidth,dimen=middle](-5.000000,-4.000000)(5.000000,-0.500000)

{ \small 
\rput[b](0.000000,-4.388889){
\begin{tabular}{c}
SNR (dB)\\
\end{tabular}
}

\rput[t]{90}(-6.3,-2.250000){
\begin{tabular}{c}
$\log_{10}\,P_\mathrm{E}$\\
\end{tabular}
}
} 

\newrgbcolor{color263.0016}{0  0  0}
\psline[plotstyle=line,linejoin=1,showpoints=true,dotstyle=Bo,dotsize=\MarkerSize,linestyle=solid,linewidth=\LineWidth,linecolor=color263.0016]
(-5.000000,-0.706646)(-4.000000,-0.771935)(-3.000000,-0.849457)(-2.000000,-0.941850)(-1.000000,-1.052403)
(0.000000,-1.185152)(1.000000,-1.345079)(2.000000,-1.538365)(3.000000,-1.772841)(4.000000,-2.057949)
(5.000000,-2.405739)

\newrgbcolor{color264.0011}{0  0  0}
\psline[plotstyle=line,linejoin=1,showpoints=true,dotstyle=B|,dotsize=\MarkerSize,linestyle=solid,linewidth=\LineWidth,linecolor=color264.0011]
(-5.000000,-0.706488)(-4.000000,-0.771753)(-3.000000,-0.849236)(-2.000000,-0.941560)(-1.000000,-1.051987)
(0.000000,-1.185152)(1.000000,-1.344527)(2.000000,-1.538019)(3.000000,-1.772700)(4.000000,-2.057950)
(5.000000,-2.405462)

\newrgbcolor{color265.0011}{0  0  0}
\psline[plotstyle=line,linejoin=1,showpoints=true,dotstyle=Bsquare,dotsize=\MarkerSize,linestyle=solid,linewidth=\LineWidth,linecolor=color265.0011]
(-5.000000,-0.762888)(-4.000000,-0.838849)(-3.000000,-0.929358)(-2.000000,-1.037638)(-1.000000,-1.167783)
(0.000000,-1.324520)(1.000000,-1.514910)(2.000000,-1.732132)(3.000000,-1.996012)(4.000000,-2.370102)
(5.000000,-2.790373)

\newrgbcolor{color266.0011}{0  0  0}
\psline[plotstyle=line,linejoin=1,showpoints=true,dotstyle=*,dotsize=0.9\MarkerSize,linestyle=solid,linewidth=\LineWidth,linecolor=color266.0011]
(-5.000000,-0.760284)(-4.000000,-0.836173)(-3.000000,-0.926289)(-2.000000,-1.034142)(-1.000000,-1.163713)
(0.000000,-1.319236)(1.000000,-1.509517)(2.000000,-1.739414)(3.000000,-2.019361)(4.000000,-2.361157)
(5.000000,-2.779459)

\newrgbcolor{color267.0011}{0  0  0}
\psline[plotstyle=line,linejoin=1,linestyle=solid,linewidth=\LineWidth,linecolor=color267.0011]
(-5.000000,-0.884850)(-4.000000,-0.985103)(-3.000000,-1.105664)(-2.000000,-1.251277)(-1.000000,-1.427899)
(0.000000,-1.643016)(1.000000,-1.906053)(2.000000,-2.228878)(3.000000,-2.626449)(4.000000,-3.117615)
(5.000000,-3.726138)

{ \small 
\rput(-1.7,-3.3){%
\psshadowbox[framesep=0pt,linewidth=\AxesLineWidth,shadowsize=0pt]
{\psframebox*{\begin{tabular}{l}
\Rnode{a1}{\hspace*{0.0ex}} \hspace*{0.7cm} \Rnode{a2}{~~Tay {\it{et al.}}'s method $(1, 1)$} \\
\Rnode{a3}{\hspace*{0.0ex}} \hspace*{0.7cm} \Rnode{a4}{~~Proposed method $(1, 1)$} \\
\Rnode{a5}{\hspace*{0.0ex}} \hspace*{0.7cm} \Rnode{a6}{~~Proposed method $(2, 1)$} \\
\Rnode{a7}{\hspace*{0.0ex}} \hspace*{0.7cm} \Rnode{a8}{~~Proposed method $(1, 2)$} \\
\Rnode{a9}{\hspace*{0.0ex}} \hspace*{0.7cm} \Rnode{a10}{~~Optimum linear detector} \\
\end{tabular}}
\ncline[linestyle=solid,linewidth=\LineWidth,linecolor=color263.0016]{a1}{a2} \ncput{\psdot[dotstyle=Bo,dotsize=\MarkerSize,linecolor=color263.0016]}
\ncline[linestyle=solid,linewidth=\LineWidth,linecolor=color264.0011]{a3}{a4} \ncput{\psdot[dotstyle=B|,dotsize=\MarkerSize,linecolor=color264.0011]}
\ncline[linestyle=solid,linewidth=\LineWidth,linecolor=color265.0011]{a5}{a6} \ncput{\psdot[dotstyle=Bsquare,dotsize=\MarkerSize,linecolor=color265.0011]}
\ncline[linestyle=solid,linewidth=\LineWidth,linecolor=color266.0011]{a7}{a8} \ncput{\psdot[dotstyle=*,dotsize=\MarkerSize,linecolor=color266.0011]}
\ncline[linestyle=solid,linewidth=\LineWidth,linecolor=color267.0011]{a9}{a10} 
}%
}%
} 

\end{pspicture}%
\centering
\caption{Error probability performance of the designed $2$-symmetric $2$-uniform tree network for different rate pairs $(R_l, R_r)$, and for different per channel SNRs.}
\label{fig:comp1}
\end{figure}
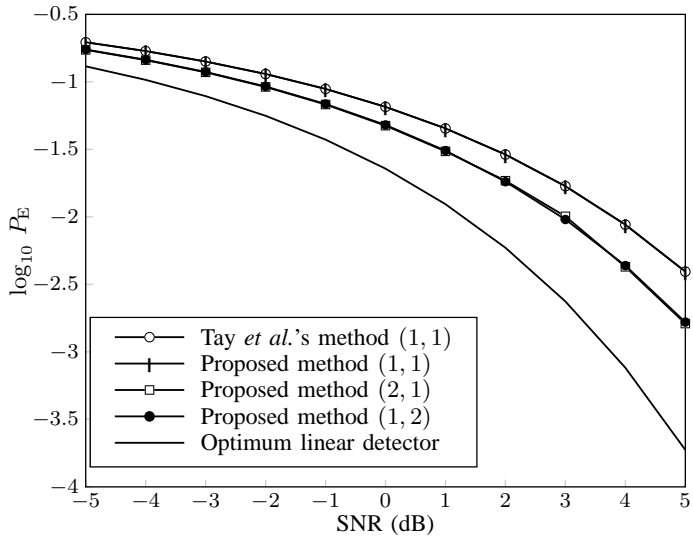
A performance comparison of the designed tree networks for different rate pairs $(R_l, R_r)$ and for different per channel SNRs is illustrated in Fig.~\ref{fig:comp1}. The results of the proposed method are compared to the optimum un-constrained linear detector (which is optimum for this problem) applied to the set of all inputs and results due to Tay \emph{et al.}'s method which leads to the optimal error exponent for an $r$-symmetric tree \cite{TTW09} for rate pair $(1, 1)$. In that case, the relays use an AND strategy and the leaves have the same threshold $\gamma$ on their observations. Using an exhaustive search, we found the best $\gamma$ which minimizes the error probability at the FC, given that the FC uses the MAP rule. The simulation results in Fig.~\ref{fig:comp1} show that for rate pair $(1,1)$ the proposed method gives the same result as the asymptotically optimum solution, and increasing the rate of the links gives better performance. Also, note that the performance of designed tree networks for rate pair $(1,2)$ coincides with that for rate pair $(2,1)$ for equally probable hypotheses. It should however be mentioned that it is not a general result and for other a-prior probability assignments, the resulting curves do not show the same performance.

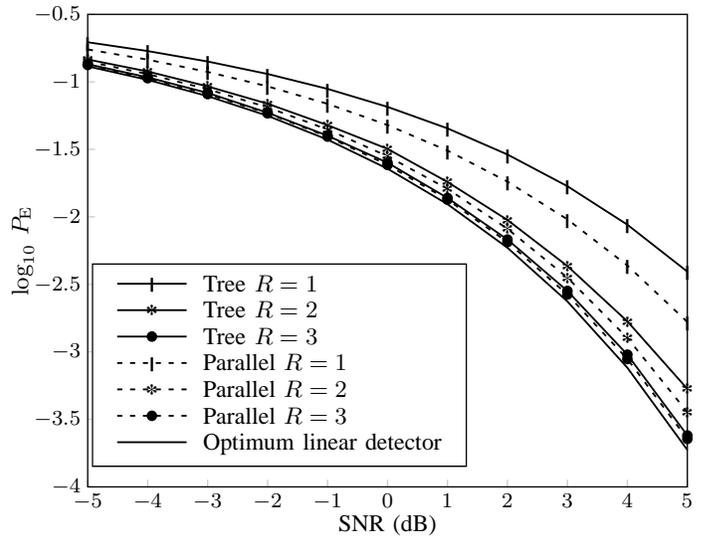
\begin{figure}[t]
%
\psset{xunit=0.100000\plotwidth,yunit=0.225346\plotwidth}%
\begin{pspicture}(-6.198157,-4.388889)(5.115207,-0.418129)%


\psline[linewidth=\AxesLineWidth,linecolor=GridColor](-5.000000,-4.000000)(-5.000000,-3.946748)
\psline[linewidth=\AxesLineWidth,linecolor=GridColor](-4.000000,-4.000000)(-4.000000,-3.946748)
\psline[linewidth=\AxesLineWidth,linecolor=GridColor](-3.000000,-4.000000)(-3.000000,-3.946748)
\psline[linewidth=\AxesLineWidth,linecolor=GridColor](-2.000000,-4.000000)(-2.000000,-3.946748)
\psline[linewidth=\AxesLineWidth,linecolor=GridColor](-1.000000,-4.000000)(-1.000000,-3.946748)
\psline[linewidth=\AxesLineWidth,linecolor=GridColor](0.000000,-4.000000)(0.000000,-3.946748)
\psline[linewidth=\AxesLineWidth,linecolor=GridColor](1.000000,-4.000000)(1.000000,-3.946748)
\psline[linewidth=\AxesLineWidth,linecolor=GridColor](2.000000,-4.000000)(2.000000,-3.946748)
\psline[linewidth=\AxesLineWidth,linecolor=GridColor](3.000000,-4.000000)(3.000000,-3.946748)
\psline[linewidth=\AxesLineWidth,linecolor=GridColor](4.000000,-4.000000)(4.000000,-3.946748)
\psline[linewidth=\AxesLineWidth,linecolor=GridColor](5.000000,-4.000000)(5.000000,-3.946748)
\psline[linewidth=\AxesLineWidth,linecolor=GridColor](-5.000000,-4.000000)(-4.880000,-4.000000)
\psline[linewidth=\AxesLineWidth,linecolor=GridColor](-5.000000,-3.500000)(-4.880000,-3.500000)
\psline[linewidth=\AxesLineWidth,linecolor=GridColor](-5.000000,-3.000000)(-4.880000,-3.000000)
\psline[linewidth=\AxesLineWidth,linecolor=GridColor](-5.000000,-2.500000)(-4.880000,-2.500000)
\psline[linewidth=\AxesLineWidth,linecolor=GridColor](-5.000000,-2.000000)(-4.880000,-2.000000)
\psline[linewidth=\AxesLineWidth,linecolor=GridColor](-5.000000,-1.500000)(-4.880000,-1.500000)
\psline[linewidth=\AxesLineWidth,linecolor=GridColor](-5.000000,-1.000000)(-4.880000,-1.000000)
\psline[linewidth=\AxesLineWidth,linecolor=GridColor](-5.000000,-0.500000)(-4.880000,-0.500000)

{ \footnotesize 
\rput[t](-5.000000,-4.053252){$-5$}
\rput[t](-4.000000,-4.053252){$-4$}
\rput[t](-3.000000,-4.053252){$-3$}
\rput[t](-2.000000,-4.053252){$-2$}
\rput[t](-1.000000,-4.053252){$-1$}
\rput[t](0.000000,-4.053252){$0$}
\rput[t](1.000000,-4.053252){$1$}
\rput[t](2.000000,-4.053252){$2$}
\rput[t](3.000000,-4.053252){$3$}
\rput[t](4.000000,-4.053252){$4$}
\rput[t](5.000000,-4.053252){$5$}
\rput[r](-5.120000,-4.000000){$-4$}
\rput[r](-5.120000,-3.500000){$-3.5$}
\rput[r](-5.120000,-3.000000){$-3$}
\rput[r](-5.120000,-2.500000){$-2.5$}
\rput[r](-5.120000,-2.000000){$-2$}
\rput[r](-5.120000,-1.500000){$-1.5$}
\rput[r](-5.120000,-1.000000){$-1$}
\rput[r](-5.120000,-0.500000){$-0.5$}
} 

\psframe[linewidth=\AxesLineWidth,dimen=middle](-5.000000,-4.000000)(5.000000,-0.500000)

{ \small 
\rput[b](0.000000,-4.388889){
\begin{tabular}{c}
SNR (dB)\\
\end{tabular}
}

\rput[t]{90}(-6.3,-2.250000){
\begin{tabular}{c}
$\log_{10}\,P_\mathrm{E}$\\
\end{tabular}
}
} 

\newrgbcolor{color271.0023}{0  0  0}
\psline[plotstyle=line,linejoin=1,showpoints=true,dotstyle=B|,dotsize=\MarkerSize,linestyle=solid,linewidth=\LineWidth,linecolor=color271.0023]
(-5.000000,-0.706488)(-4.000000,-0.771753)(-3.000000,-0.849236)(-2.000000,-0.941560)(-1.000000,-1.051987)
(0.000000,-1.185152)(1.000000,-1.344527)(2.000000,-1.538019)(3.000000,-1.772700)(4.000000,-2.057950)
(5.000000,-2.405462)

\newrgbcolor{color272.0018}{0  0  0}
\psline[plotstyle=line,linejoin=1,showpoints=true,dotstyle=Basterisk,dotsize=\MarkerSize,linestyle=solid,linewidth=\LineWidth,linecolor=color272.0018]
(-5.000000,-0.834002)(-4.000000,-0.922896)(-3.000000,-1.033636)(-2.000000,-1.161864)(-1.000000,-1.320658)
(0.000000,-1.496355)(1.000000,-1.741433)(2.000000,-2.022885)(3.000000,-2.364329)(4.000000,-2.773922)
(5.000000,-3.273950)

\newrgbcolor{color273.0018}{0  0  0}
\psline[plotstyle=line,linejoin=1,showpoints=true,dotstyle=*,dotsize=\MarkerSize,linestyle=solid,linewidth=\LineWidth,linecolor=color273.0018]
(-5.000000,-0.869002)(-4.000000,-0.966401)(-3.000000,-1.081995)(-2.000000,-1.228440)(-1.000000,-1.398899)
(0.000000,-1.598702)(1.000000,-1.859961)(2.000000,-2.167893)(3.000000,-2.548368)(4.000000,-3.020457)
(5.000000,-3.618942)

\newrgbcolor{color274.0018}{0  0  0}
\psline[plotstyle=line,linejoin=1,showpoints=true,dotstyle=B|,dotsize=\MarkerSize,linestyle=dashed,dash=2pt 3pt,linewidth=\LineWidth,linecolor=color274.0018]
(-5.000000,-0.759984)(-4.000000,-0.835440)(-3.000000,-0.926289)(-2.000000,-1.033198)(-1.000000,-1.162720)
(0.000000,-1.319947)(1.000000,-1.509083)(2.000000,-1.738862)(3.000000,-2.018653)(4.000000,-2.360244)
(5.000000,-2.778275)

\newrgbcolor{color275.0018}{0  0  0}
\psline[plotstyle=line,linejoin=1,showpoints=true,dotstyle=Basterisk,dotsize=\MarkerSize,linestyle=dashed,dash=2pt 3pt,linewidth=\LineWidth,linecolor=color275.0018]
(-5.000000,-0.849356)(-4.000000,-0.942388)(-3.000000,-1.054084)(-2.000000,-1.188608)(-1.000000,-1.352813)
(0.000000,-1.550857)(1.000000,-1.792348)(2.000000,-2.088160)(3.000000,-2.449867)(4.000000,-2.896448)
(5.000000,-3.447000)

\newrgbcolor{color276.0018}{0  0  0}
\psline[plotstyle=line,linejoin=1,showpoints=true,dotstyle=*,dotsize=\MarkerSize,linestyle=dashed,dash=2pt 3pt,linewidth=\LineWidth,linecolor=color276.0018]
(-5.000000,-0.874615)(-4.000000,-0.972793)(-3.000000,-1.090753)(-2.000000,-1.233076)(-1.000000,-1.405495)
(0.000000,-1.615195)(1.000000,-1.872378)(2.000000,-2.186713)(3.000000,-2.575065)(4.000000,-3.052970)
(5.000000,-3.644803)

\newrgbcolor{color277.0018}{0  0  0}
\psline[plotstyle=line,linejoin=1,linestyle=solid,linewidth=\LineWidth,linecolor=color277.0018]
(-5.000000,-0.884850)(-4.000000,-0.985103)(-3.000000,-1.105664)(-2.000000,-1.251277)(-1.000000,-1.427899)
(0.000000,-1.643016)(1.000000,-1.906053)(2.000000,-2.228878)(3.000000,-2.626449)(4.000000,-3.117615)
(5.000000,-3.726138)

{ \small 
\rput(-1.8,-3.1){%
\psshadowbox[framesep=0pt,linewidth=\AxesLineWidth,shadowsize=0pt]{\psframebox*{\begin{tabular}{l}
\Rnode{a1}{\hspace*{0.0ex}} \hspace*{0.7cm} \Rnode{a2}{~~Tree $R=1$} \\
\Rnode{a3}{\hspace*{0.0ex}} \hspace*{0.7cm} \Rnode{a4}{~~Tree $R=2$} \\
\Rnode{a5}{\hspace*{0.0ex}} \hspace*{0.7cm} \Rnode{a6}{~~Tree $R=3$} \\
\Rnode{a7}{\hspace*{0.0ex}} \hspace*{0.7cm} \Rnode{a8}{~~Parallel $R=1$} \\
\Rnode{a9}{\hspace*{0.0ex}} \hspace*{0.7cm} \Rnode{a10}{~~Parallel $R=2$} \\
\Rnode{a11}{\hspace*{0.0ex}} \hspace*{0.7cm} \Rnode{a12}{~~Parallel $R=3$} \\
\Rnode{a13}{\hspace*{0.0ex}} \hspace*{0.7cm} \Rnode{a14}{~~Optimum linear detector} \\
\end{tabular}}
\ncline[linestyle=solid,linewidth=\LineWidth,linecolor=color271.0023]{a1}{a2} \ncput{\psdot[dotstyle=B|,dotsize=\MarkerSize,linecolor=color271.0023]}
\ncline[linestyle=solid,linewidth=\LineWidth,linecolor=color272.0018]{a3}{a4} \ncput{\psdot[dotstyle=Basterisk,dotsize=\MarkerSize,linecolor=color272.0018]}
\ncline[linestyle=solid,linewidth=\LineWidth,linecolor=color273.0018]{a5}{a6} \ncput{\psdot[dotstyle=*,dotsize=\MarkerSize,linecolor=color273.0018]}
\ncline[linestyle=dashed,dash=2pt 3pt,linewidth=\LineWidth,linecolor=color274.0018]{a7}{a8} \ncput{\psdot[dotstyle=B|,dotsize=\MarkerSize,linecolor=color274.0018]}
\ncline[linestyle=dashed,dash=2pt 3pt,linewidth=\LineWidth,linecolor=color275.0018]{a9}{a10} \ncput{\psdot[dotstyle=Basterisk,dotsize=\MarkerSize,linecolor=color275.0018]}
\ncline[linestyle=dashed,dash=2pt 3pt,linewidth=\LineWidth,linecolor=color276.0018]{a11}{a12} \ncput{\psdot[dotstyle=*,dotsize=\MarkerSize,linecolor=color276.0018]}
\ncline[linestyle=solid,linewidth=\LineWidth,linecolor=color277.0018]{a13}{a14}
}%
}%
} 

\end{pspicture}%
\centering
\caption{Comparison of error probability performance of the designed tree network with rate pairs $(R, R)$ and parallel network with rate $R$.}
\label{fig:comp2}
\end{figure}

The proposed method for the design of general tree network can also be used for the design of parallel networks. It is a well known statement that the performance of any optimum tree network is dominated by the performance of an optimum parallel network, for an equal number of observations \cite{TTW08a}. Fig.~\ref{fig:comp2} shows the error probability performance of designed tree and parallel networks, where the rate of all the links in the tree network and the rate of all the links in the parallel network are equal to $R$. As is illustrated in Fig.~\ref{fig:comp2}, for the same channel rates, the parallel network outperforms the corresponding tree network. As the rate of the links increases, the performance of the tree network and the parallel network converges to the un-constrained ($R=\infty$) case and the performance of tree network will asymptotically be the same as parallel network.

We can also see that the simulation results of the proposed numerical method are in line with what can be expected in situations where the optimal decision functions are obvious. For example, consider a $2$-symmetric $2$-uniform tree network with rate pair $(1,2)$, where the leaves send one-bit messages to the relays and the relays send two-bit messages to the FC. In this case, an optimal relay would simply put the one-bit received messages from its predecessors together and send the resulting two-bit message to the FC, which means the performance of the (optimal) $(1,2)$ tree network is the same as the performance the (optimal) parallel network with one-bit channel rates. This is consistent with Fig.~\ref{fig:comp1} and Fig.~\ref{fig:comp2}, where using the proposed design method yields the same performance in terms of error probability for both cases, which indicates that the proposed method is working as expected.

\section{Conclusion}\label{sec:conclusion}
In this paper, we have considered the distributed hypothesis testing problem in a general tree network where the nodes make observations which are, conditioned on the true hypothesis, independent. We have shown that the design of nodes under the \pbp methodology is analogous to the design of a two-node network, the \emph{restricted} model, in which the decision function can be designed efficiently (cf. \cite{Alla14Tsp}). We also have shown how the parameters of the restricted model for the design of a node in the general tree can be formed in a recursive and computationally efficient manner.

\section*{Acknowledgment}
This work has been supported in part by the ACCESS seed project DeWiNe.

\bibliographystyle{IEEEtran}
\bibliography{Ref}

\end{document}